\documentclass[a4paper]{jpconf}
\usepackage{graphicx}
\begin{document}
\title{The Origin of Parsec-Scale Gaseous  and Stellar Disks in the Galactic Center and AGNs}

\author{F. Yusef-Zadeh}
\address{Dept. Physics and Astronomy, Northwestern University, Evanston, IL. 60208}
\ead{zadeh@northwestern.edu}

\author{M. Wardle}
\address{Department of Physics and Astronomy, Macquarie University, Sydney, NSW 2109, Australia}
\ead{mark.wardle@mq.edu.au}

\begin{abstract} The Galactic center stellar disk and the circumnuclear ring provide a unique 
opportunity  to study in detail the dynamics and physical conditions of 
distant molecular disks in the nuclei of galaxies. 
One of the key 
questions is how these disks form  so close to their host black holes and 
under what condition they form stars in a tidally stressed environment. 
We argue that disk formation around a massive black hole is 
due to partial accretion of extended molecular clouds 
that temporarily pass through the central region of the Galaxy. The 
cancellation of angular momentum of the gravitationally focused 
gas naturally creates a compact gaseous disk. 
The disk  can potentially become 
gravitationally  unstable and form stars. We apply these ideas to explain the origin 
of sub-parsec  megamaser disks found in the nuclei of Seyfert 2 galaxies. 
We show that  an empirical  scaling relation  between the mass of 
the black hole and the size of the disk  can be understood in the context 
of the cloud capture scenario. 
We conclude that the stellar and gas disks found in our Galactic center 
act  as a bridge to further our understanding of more 
distant mega-maser disks 
in the nuclei of Seyfert 2 galaxies.

\end{abstract}


\section{Introduction}

Understanding the formation of massive young stars in the immediate 
vicinity of massive black holes is challenging. Two plausible 
models for  the origin of the disk of massive stars 
orbiting Sgr A*, the 4$\times10^6$ M$_{\odot}$ black hole
at the Galactic center, 
have been proposed, 
namely, in-situ star formation and the dynamical 
migration of star clusters (see Genzel, Eisenhauer and Gillessen 
2010; Levin and Beloborodov 
2003; Gerhard 2001; Nayakshin et al. 2007). It has become clear that the 
in-situ star formation model can account for present observations better than 
the migration scenario in which a young star cluster undergoes 
dynamical friction and spirals  toward the Galactic center. The question then 
arises as to what extent disk formation near Sgr A* can 
provide insight in the formation of sub-parsec gaseous disks orbiting 
supermassive black holes in AGNs. To address this question, we will first 
focus on the origin of Galactic center stellar and molecular disks near 
Sgr A* followed by AGN disks in the context of partial cloud capture by 
the massive black hole. We then discuss observational signatures 
such as the scaling relation between the mass of the black hole and the size of
the disk as well as 
the recoil of the black 
hole 
in the context of the cloud capture  scenario.  A more detailed account of the 
results presented here can be found in Wardle and Yusef-Zadeh (2008, Paper 
I), Yusef-Zadeh et al. (2009, Paper II) and Wardle and Yusef-Zadeh (2011, 
Paper III).

\section{Galactic Center Young Stellar Disks} 

Observations of our Galactic center indicate that there are about 100 massive young stars 
distributed 
within one or two disks  with a sharp inner edge (Paumard et al. 2006; Lu et al. 2010).  
In the in-situ formation scenario, an interstellar cloud is tidally disrupted and captured by Sgr A* (Nayakshin, 
Cuadra \& Spingel 2007). 
There are two issues   that are of interest in the context of 
this mode  of star formation. 
One  is  related to 
the mechanism in  which  massive stars are formed in a disk. 
If the gas disk is significantly massive, it can become 
gravitationally 
unstable 
resulting in the fragmentation and formation of stars. 
A cooling self-gravitating gas disk around a black hole becomes unstable to 
fragmentation on a dynamical time scale if
Toomre's  Q$<1$. 
In this picture,  the strong  tidal force by the black hole plays a dual
role. On the one hand, 
tidal shear acting  against self-gravity 
stretches  the cloud in the radial direction, making it 
more difficult 
for the cloud to fragment. On the other hand, 
the gas is tidally squeezed in  the vertical direction, 
parallel to the disk axis, 
thus,  making it easier to resists the gravitational stress by 
increasing  the gas density to the Roche's critical value. 
As the cooling time is comparable to the dynamical time scale, stars in a hot disk 
are formed in self-gravitating  clumps 
with a range of eccentricities and 
inclinations as the disk circularizes (Nayakshin et al. 2007; 
Wardle and Yusef-Zadeh 2008; Bonnel and Rice 2008; Alig et al. 2011). 
Simulations 
of the capture of a cloud
suggests that star formation during such an
event will occur before the disk is  fully circularized and 
becomes dynamically cold (Mapelli et al.\ 2008). 
This is in contrast to a cold self-gravitating disk that
forms stars with  small eccentricities. 
Large eccentricities could not easily be built up from initial circular orbits in a
cold disk, thus, the hot disk scenario may better explain the observed
stellar kinematics

The second  issue  is related the mechanism in which a cloud loses angular momentum 
to form a disk  so close to Sgr A*.  The cloud has no means of getting rid of its 
angular momentum and it is difficult and rare to capture a 
compact cloud by a massive black hole  without being engulfed. 
However, nature provides an easier mechanism to bring the 
gaseous material closer in to the black hole. This is done 
 by the capture of material from an extended cloud passing  on opposite sides of 
the black hole, thus allowing 
the loss of much of the  angular momentum  of gravitationally 
focused gas. 
The gaseous material with large impact parameters can 
settle much closer to the black hole, thus higher gas density is achieved. 
Given  the level of asymmetry and  inhomogeneity in the spatial  distribution 
of  the cloud, 
the cloud capture event may even account for 
multiple 
counter-rotating disks  with different inclinations.
Figure 1 shows a diagram of a cloud impacting Sgr A*. The gravitational focusing of an incoming 
molecular cloud is shown in the top panel whereas the outer  region of the cloud that is 
not captured continuous its motion in the direction away from Sgr A*. 
The inner region of the cloud is carved out, captured and circularized to form a disk.

\section{Galactic Center Molecular Disk}

Multi-wavelength studies of the
Galactic center  show that
a 2-parsec clumpy molecular ring 
circles Sgr A*  with a rotational
velocity of about 100 km s$^{-1}$ (Christopher et al. 2005). 
The observed CO  and HCN (G\"usten et al.\ 1987) emission indicates
a total mass of $\sim 10^5$ 
M$_{\odot}$.  However, 
HCN 
observations imply that there is high density material close to the inner 
edge (Christopher  et al. 2005) suggesting that the 
mass may be closer to $10^6$ M$_{\odot}$.  
The application of the formation  scenario to the
circumnuclear ring of gas argues that this disk was recently 
captured, is currently  settling down  and is on
the verge of forming stars (Paper I, Paper III).

The extent of the ring suggests an initial migrating cloud speed towards 
the lower end of the 50--100 km s$^{-1}$  range. 
The Hoyle-Lyttleton 
radius is about 12\, pc which is  
similar to  the  outer radius of the molecular ring. 
 On this scale, one expects 
considerable asymmetry in the cloud material passing by Sgr A* during the 
capture event, with a corresponding reduction in the net cancellation of 
angular momentum during the capture process (cf.\ Bottema \& Sanders 1986).  
At first sight our model suggests that the circumnuclear ring could  become
 unstable to gravitational fragmentation (Paper I), 
but this assumes that it 
has kinematically relaxed.  The velocity dispersion of the ring is $\sim$30 km 
s$^{-1}$, so that it is not unstable unless its mass is $10^6$\,M$_{\odot}$.

Methanol and 
water masers have recently been detected 
suggesting that  star formation is taking place
(Paper II).  It 
appears that the circumnuclear ring is still in the process of settling 
down soon after formation.  The ring's orbital time scale at 2 pc is $\sim 
10^5$\,yr, so this implies that the age of the ring is $\leq 10^6$\,yr.  
If this is so, the remains of the original interloper cloud should lie 
within $\sim 100$\,pc of Sgr A*. One candidate is the +50 km s$^{-1}$\ molecular 
cloud which extends along the plane from the Galactic center to 
l$\approx0.2^0$ and consists of a number of bound cloudlets with a total 
mass of $\sim10^6$ M$_{\odot}$. This cloud is 
thought to lie about 30\,pc behind Sgr A*, consistent with an interaction 
$\sim3\times10^5 $ years ago.

\section{AGN Megamaser Disks}

Recent VLBA observations have discovered several sub-parsec  megamaser disks with high inclinations 
in the nuclei of Seyfert 2 
galaxies (Herrnstein et al. 2008; Kuo et al. 2010). 
The inner and outer radii are determined by milli-arcsecond observations. 
The mass of the disks are measured accurately
because these  
disks lie well within the sphere of influence of their host black holes 
and are not self-gravitating. 
Figure 2 shows an 
empirical linear correlation between the disk radius and the black hole mass. 
There is a well-defined upper envelope to the maser disk radii which scales linearly with black hole mass, given
approximately by                 
    $\rm R_\mathrm{max} \sim 0.3\,\,M_7\;\mathrm{pc} \,, $ 
where the central black hole mass is
$M=M_7\times10^7$\,M$_\odot$.

The remarkable similarity of size scale of the stellar disk surrounding Sgr A* and the megamaser 
disks suggest that they could be formed the same way. Paper III shows that  for plausible 
estimates of 
the mass and angular momentum of the captured material, this process naturally reproduces the 
empirical linear relationship between maser disk size and black hole mass.  The capture of clouds 
with column densities $\leq 10^{23.5}$ cm$^{-2}$ results in a non-self-gravitating disks of the 
correct scale and sufficient column density to allow X-ray irradiation from the central source to 
reproduce thin megamaser disks.  By contrast, the capture of a cloud with higher column density 
would instead create a self-gravitating disk giving rise to rapid star formation.

In the context of disk formation, the disk size scales with black hole 
mass. The reason for such a scaling is as follows. The mass of the disk 
depends on the Hoyle-Littelton capture radius and the column density of 
the initial cloud. The size of the disk depends on the capture radius 
which itself depends on the mass of the black hole and the angular 
momentum cancellation. Thus, the cloud capture model implies 
that the radius of the disk depends 
linearly on 
the mass of the black hole.

\section{Cloud Migration}

How do molecular  clouds migrate toward Galactic nuclei? 
The inner 200\,pc of the Galaxy is rich in dense molecular clouds, many of which 
are on eccentric orbits (Bally et al.\ 1988; Oka et al.\ 1998; Martin et al.\ 2004; Jones et al. 2011).  In addition to 
the +50 km 
s$^{-1}$ molecular cloud and other members of a disk population of molecular clouds distributed within the inner 
30\,pc of Galactic center.  Their non-circular, elongated motion is thought to be induced by the Galaxy's barred 
potential (see Kim et al. 2011), with dynamical 
friction 
aiding migration to the central regions of the Galaxy. 
The bar potential of the 
Galaxy 
can apply  a torque to the clouds, thus trigger infall
toward the nucleus (see F. Combes in these proceedings).  
Recent simulations also suggest that gas supplied to galactic 
centers is controlled by angular momentum 
transfer from one massive gas clump to another during gravitational encounters (Namekata \& Habe 2011).
In the context of a transition from 
X1 to X2 orbits, the gas clouds lose angular momentum spiraling in toward 
the central region of the Galaxy. In order to bring gaseous material close in to the inner
few pc,  the bar must be  nested (Namekata et al. 2009).

Some of these inward-moving clouds may interact with their host supermassive black holes.  
In our own 
Galaxy, the circumnuclear molecular ring (e.g.\ Christopher et al. 2005) at a distance of 1.7\,pc 
from Sgr A*, the 50 km s$^{-1}$ molecular cloud,  
and the circumnuclear rings found on scales of several parsecs from 
the center of 
numerous Seyfert galaxies suggest an ample supply of material.  
The rate of
migration of molecular material is estimated to be about $10^4-10^5$ {\hbox{M$_\odot$}}
per $10^6-10^7$ years.
The age of the stellar disk  (Paumard et al.\ 2006), and the relative youth 
of the 
circumnuclear ring (Paper II) is consistent with the age estimate  implying 
that the rate of encounters of 
massive clouds with Sgr A* is 
$\sim 10^{-6}\ $ yr$^{-1}$.  
This may have been been ongoing for a significant fraction of the Galaxy's lifetime as the stellar population in 
the central parsec is consistent with roughly constant star formation over the past 12\,Gyr (Maness et al.\ 
2007). 

\section{Black Hole Recoil}

The partial capture of a cloud with incident  velocity $V_{initial}$ 
imparts momentum to the system of black hole with a mass M$_{BH}$ and gaseous 
disk with  a mass  $M_{Disk}$, 
implying    a kick velocity 
\begin{equation}
V_{kick} = \frac{M_{Disk}\times\,  V_{initial}} {M_{BH} +  M_{Disk}}
\end{equation}
to the black hole. 
Given that black holes like Sgr A* are embedded in their host  nuclear cluster, the black hole 
is expected to move with respect to the center of the cluster 
and exchanges heat with the cluster due to dynamical friction.  
The recoil is of the order of few km s$^{-1}$ and is 
likely to be damped via  dynamical friction on the 
surrounding stars, on a 
timescale of the  stellar crossing time across the sphere of influence, $\sim 10^3$\,years.
However, the parameter space for  the significance of the recoil 
to the black hole in different 
nuclear cluster environment  has not been explored. 

\section{Conclusions }

In summary, we have investigated the capture  of  gas clouds by massive black holes via
the Hoyle-Lyttleton  process.
We applied this mechanism to three different evolutionary phases of cloud capture scenario. 
One is a subset of AGN megamaser disks with high inclinations.  There is 
no apparent star formation in these 
disks which lie  within the sphere of influence of their host black holes.
Second  application of the cloud capture scenario 
is to the  2-parsec molecular disk orbiting Sgr A* where the gaseous disk  is 
young. It has been argued that  this system is 
showing  early  signatures of star formation. Lastly, 
sub-parsec stellar disk orbiting Sgr A* where massive star formation 
has already taken  place a
few million years ago.

The lack of star formation in  megamaser disk  systems could be due to 
strong magnetic fields acting against self-gravity of the disk  
(Johansen and Levin 2008; Gaburov, Johansen and Levin 2011). 
Alternatively, the low column density may be responsible. 
Assuming that  the tidal shear by the black hole is accounted for, then the mass-to-flux ratio 
is an important parameter for the comparison of  self-gravity 
and the magnetic field.  Thus, the lack   of  star formation depends 
on the column density of incoming clouds and the magnetic field of the disks  rotating around  massive black 
holes.  Future  molecular line and magnetic field observations of 
thermal gas in these  disks  should be quite useful. 

The physical relationship that threads through these different 
systems is the capture of a cloud by
engulfing  the massive black hole. This is an effective way to 
get rid of the angular momentum of the incoming molecular cloud. 
The  engulfing scenario can also be applied when a star 
approaches the tidal disruption radius of a massive black hole. 
We speculate that these  processes  could operate in 
galaxy mergers, dense stellar cusps around massive black holes, leading 
to starburst activity, black hole growth  and powerful  flaring activity 
 in the vicinity of  central massive black holes.

We found a  scaling relation between the radius of the captured disk and the massive 
black hole which  appears to be supported 
by the Hoyle-Lyttleton  capture process. The outer radius of the captured cloud is understood
in terms of a physical truncation of the disk because the scale determined                               
kinematics of cloud capture depends linearly on the black hole mass.            
The apparent  inner radius of the captured disk  can be understood 
if the disk is warped in such a way that the inner disk does not get 
sufficient exposure to the central source of 
X-rays, to  produce 
22\,GHz H$_2$O  (Maloney et al. 2002). 
As for the  disk of stars orbiting Sgr A*, the inner radius  is 0.03pc. 
In this case, it is possible that the stars that are formed within the inner 0.03pc 
are influenced by 
dynamical processes. In fact, there is a cluster of B-type stars with eccentric orbits 
distributed isotropically within the inner radius 
but with a completely different distribution than that of the disk of stars. 
Alternatively, it is possible that  the captured gas within the inner radius 
is   dissipated to high temperature and does not have sufficient density to overcome 
the tidal stress.  This idea could be relevant to the formation of the 
inner radius of the 
circumnuclear molecular ring where  hot  stars of  the disk are thought  to have ionized the inner 1.7pc of molecular gas 
and created  a cavity of
molecular gas in this region.


\section{References}

\smallskip

\begin{figure}[h]
\begin{minipage}{18pc}
\includegraphics[width=18pc]{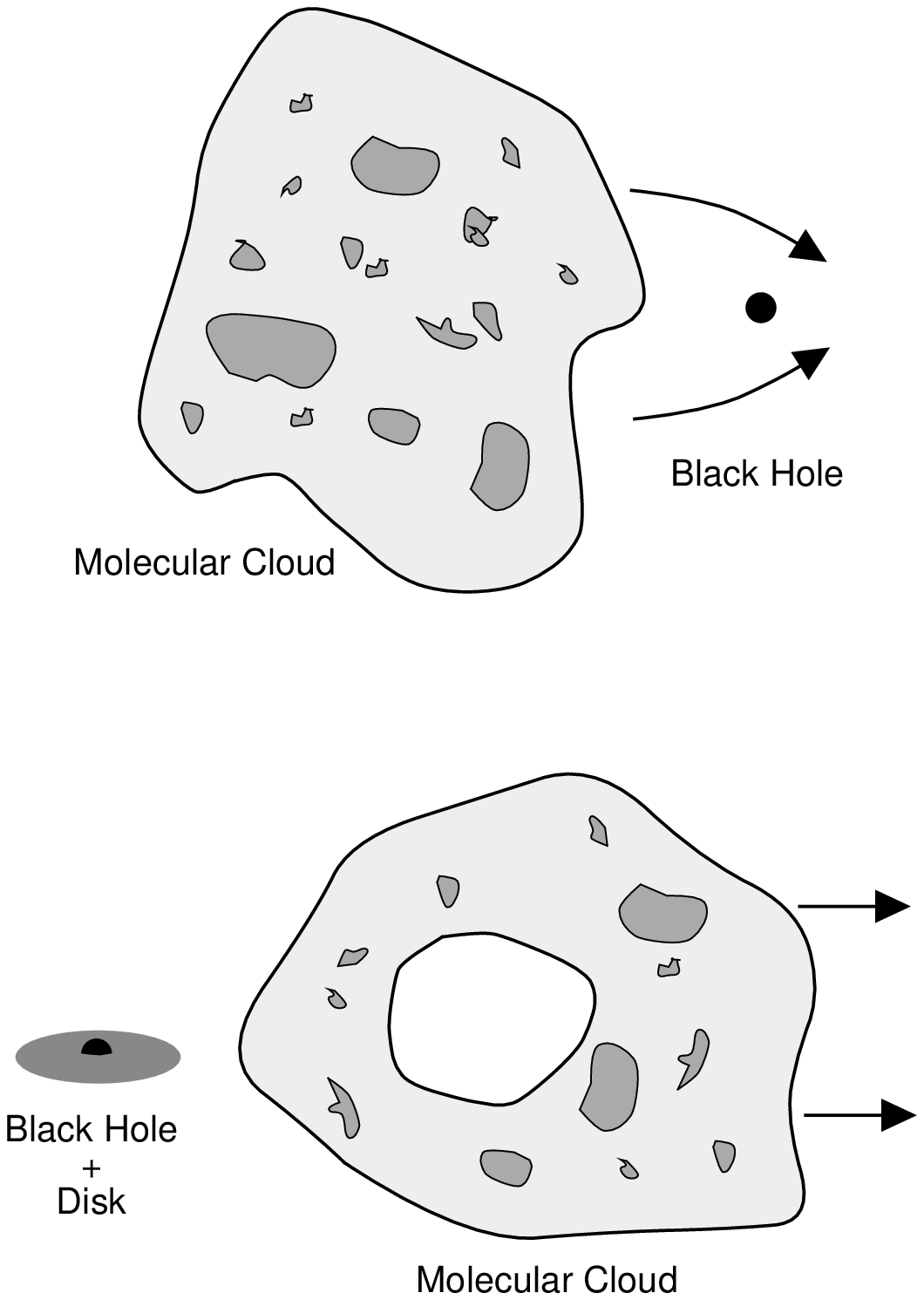}
\caption{\label{label}A schematic diagram of a cloud before and after impacting Sgr A*.  }
\end{minipage}\hspace{2pc}%
\begin{minipage}{18pc}
\includegraphics[width=18pc]{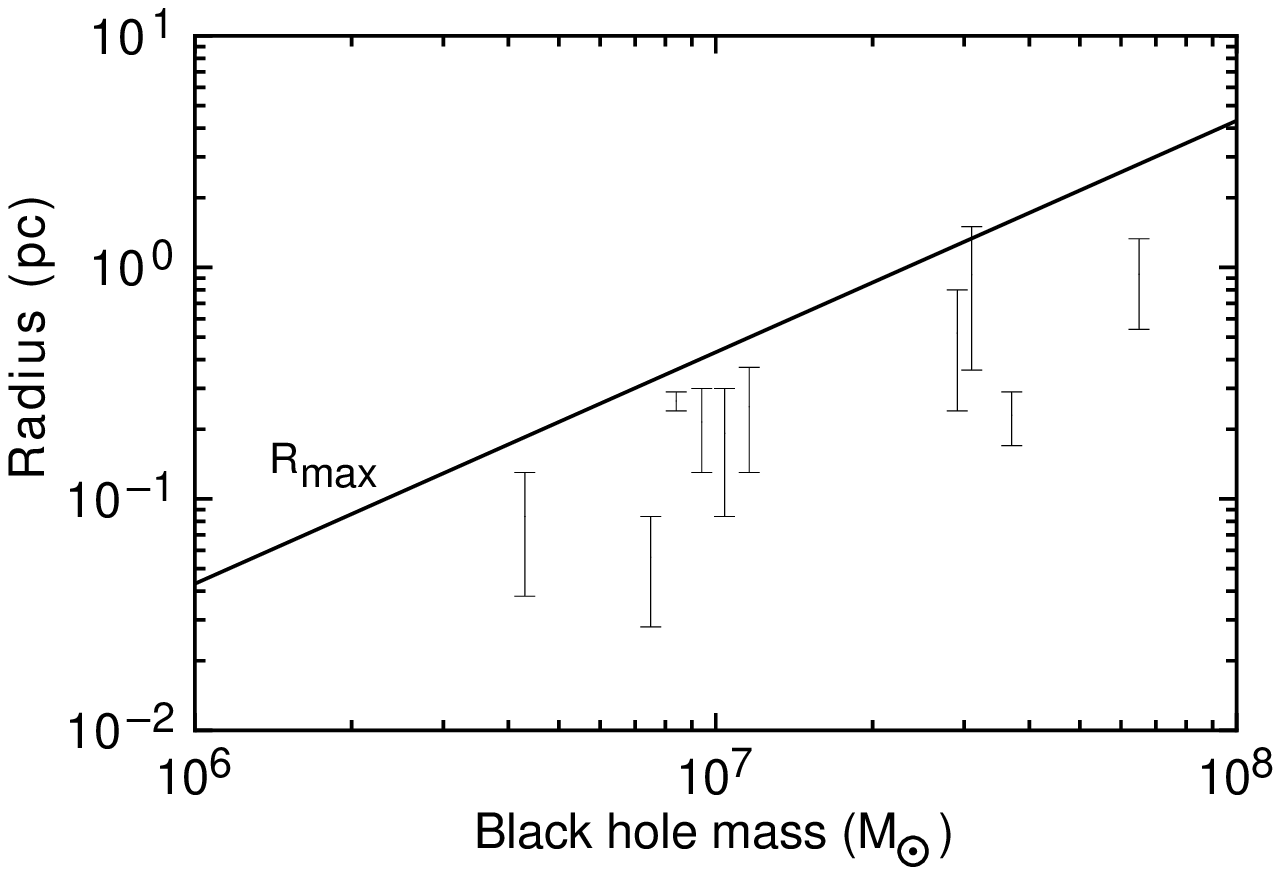}
\caption{\label{label}
A plot of the black hole mass and the outer and inner radii 
of AGN mega-maser AGNs plus  the
Galactic center stellar disk. The upper envelope, labelled ''$R_\mathrm{max}$'', is given by
$R_\mathrm{max} = 0.3 (M/10^7{\mathrm{M_\odot}})$\,pc. }
\end{minipage} 
\end{figure}

\end{document}